# Mapping the Geography of Science: Distribution Patterns and Networks of Relations among Cities and Institutes

Loet Leydesdorff[1] & Olle Persson[2]

**Abstract**

Using Google Earth, Google Maps and/or network visualization programs such as Pajek, one can overlay the network of relations among addresses in scientific publications on the geographic map. We discuss the pros en cons of the various options, and provide software (freeware) for bridging existing gaps between the *Science Citation Indices* and *Scopus*, on the one side, and these various visualization tools, on the other. At the level of city names, the global map can be drawn reliably on the basis of the available address information. At the level of the names of organizations and institutes, there are problems of unification both in the ISI-databases and *Scopus*. Pajek enables us to combine the visualization with statistical analysis, whereas the Google Maps and its derivates provide superior tools at the Internet.

[1] Amsterdam School of Communications Research (ASCoR), University of Amsterdam, Kloveniersburgwal 48, 1012 CX Amsterdam, The Netherlands; loet@leydesdorff.net.
[2] Department of Sociology, Umeå University, SE 901 87 Umeå, Sweden; Olle.Persson@soc.umu.se.

# Mapping the Geography of Science: Distribution Patterns and Networks of Relations among Cities and Institutes

## 1. Introduction

In this communication we report on newly available methodologies to map the sciences both statically (at each moment of time) and dynamically (over time). These techniques enable us, among other things, to visualize patterns of international collaboration using a projection on the world map (e.g., Glänzel, 2001; Hicks & Katz, 1996; Persson *et al.,* 2004; Wagner, 2008; Zitt *et al.*, 1999). We compare the different possibilities in Google Earth, Google Maps, and Pajek, and report on dedicated software (freeware) available for making these projections using data from bibliographic databases such as the *Science Citation Index* and *Scopus*.

The geographic mapping of science can be distinguished from its cognitive mapping (Frenken *et al*., 2009; Jones *et al*., 2008; Small & Garfield, 1985). The sciences can be mapped cognitively, for example, in terms of journal maps (e.g., Leydesdorff, 1986; Tijssen *et al*., 1987), co-citations (Small & Griffith, 1974; Small, 1999), or co-words (Callon *et al*., 1983). Using techniques such as multi-dimensional scaling (e.g., Kruskal & Wish, 1978; Borgatti, 2001; Leydesdorff & Schank, 2008) or spring-embedded algorithms (e.g., Kamada & Kawai, 1989; Fruchterman & Reingold, 1999), information scientists have made considerable advances during the last decade in terms of agreeing on similarity criteria (Ahlgren *et al*., 2003; Van Eck & Waltman, 2009), possible projections (Boyack *et al*., 2005 and 2007; De Moya-Anegón *et al*., 2007; Klavans & Boyack, 2009a; Rafols & Leydesdorff, 2009), and even on standard colors for distinguishing among disciplinary affiliations (Klavans & Boyack, 2009b). The latter authors suggest that "consensus" has emerged on the mapping. Rafols *et al*. (in preparation) drew the conclusion that therefore one would be able to project developments in science against a statistical baseline.

Since in a socio-cognitive process such as the development of the sciences, change can take place at different levels at the same time, Studer & Chubin (1982, at p. 269) already noted that "(r)elationships among journals, individuals, references, and citations can be analyzed in terms of their structural properties. But can one be used as a baseline to calibrate our understanding of another? Does it make sense to attempt to "control" for one relationship while studying others?" Narin (1976) was the first to distinguish between nations and disciplines as two analytically independent baselines in the evaluation (cf. Narin *et al*., 1972; Narin & Carpenter, 1975). Small & Garfield (1975) proposed to use these two dimensions as different bases for the mapping.

Intellectual developments at the global level have also to be retained locally. National (or regional) governments develop science and technology policies for this retainment (e.g., Skolnikoff, 1993). Does investment in science pay off in terms of prominence and reputation, economic returns, or the emergence of transnational linkages such as envisaged by the European Commission? (Leydesdorff & Wagner, 2008, 2009; NSB, 2010, pp. 5-33 ff.). Are national governments able to formulate policies which provide them with a possible hold on "emerging technologies"? Is sufficient knowledge infrastructure developed to play a role in the case of "generic technologies"? These and similar questions require a geographic baseline for the assessment in addition to the cognitive map.

The geographic map, of course, provides us with a natural baseline for studying the spatial dynamics. In recent years, software developments have made this map increasingly available for the projection at different scales and with appropriate zooming techniques, such as in Google Maps. How can one make such techniques profitable for the enterprise of science and technology studies? Having both of us been deeply involved in developing software for using the information contained in databases such as the *Science Citation Index* and *Scopus* for the mapping, we thought it timely to provide a state-of-the-art review of the current possibilities and limitations of geographic maps. Where necessary we further developed our software for bridging gaps and made these tools available from our respective websites. The interested reader can find instructional materials and manuals at these sites (http://www.leydesdorff.net/maps and http://www8.umu.se/inforsk/bibexcel/, respectively).

## 2. Methods and materials

For didactic purposes we shall use a standard set for the various visualizations. We chose to use the footprint of the field of information science (IS) in 2009 as available in the address information in the bylines of the publications. How did we delimit this field?

First, Library & Information Science (LIS) is categorized as a separate subject in the *Social Science Citation Index,* but this category covers 61 journals. These lists, however, are composed for the purpose of information retrieval and therefore not sufficiently restricted for mapping a specific field (Leydesdorff & Probst, 2009). More restricted lists of IS have been proposed in the literature. Based on White & McCain (1998, at p. 300), Zhao & Strotman (2008, at p. 2072) recently provided an updated list of eight journals representing the core of IS, in their opinion. Using aggregated journal-journal citation data from the *Journal Citation Report* 2008, we found 13 journals that contribute more than 1% to the citations of *JASIST* and 11 journals that contribute more then 1% to the citations of *Scientometrics.* These two sets overlap in eight journals, of which six are also included in the list of Zhao & Strotman (2008).

Since we wished to include also the newly added *Journal of Informetrics*, we gave priority to our citation-based definition of the field and included these eight journals in the analysis (Table 1). Using a search string based on these eight journals, 621 *articles* could be retrieved from the Web-of-Science published in the year 2009.[1] Because we limited the set to articles, however, no records from the *Annual Review of Information Science and Technology*—including 10 reviews and one editorial in 2009—were retrieved. We limited the analysis to this set (available at http://www.leydesdorff.net/maps/data.zip). The file includes 1,479 authors at 1,107 institutional addresses.

[1] The records were downloaded on January 14, 2009. The search string was: so=(annual review of information science “and” technology or journal of information science or journal of information science or journal of the american society for information science “and” technology or scientometrics or journal of informetrics or information processing management or information research an international electronic journal or journal of documentation) and document type=(article) timespan=2009

| | Zhao & Strotman (2008) | Citation environment *JASIST* (2008) | Citation environment *Scientometrics* (2008) | Journals included in this analysis |
|---|---|---|---|---|
| *ACM Transations of Information Systems* | | + | | |
| ***Annual Review of Information Science and Technology*** | + | + | + | **(+)** |
| *Computation and Human Behavior* | | + | | |
| *Decision Support Systems* | | + | | |
| ***Information Processing & Management*** | + | + | + | + |
| ***Information Research*** | | + | + | + |
| ***Journal of the American Society for Information Science and Technology*** | + | + | + | + |
| ***Journal of Documentation*** | + | + | + | + |
| ***Journal of Informetrics*** | | + | + | + |
| ***Journal of Information Science*** | + | + | + | + |
| *Library & Information Science Research* | + | + | | |
| *Knowledge Organization* | | + | | |
| *Online Information Review* | | | + | |
| *Proceedings of the ASIST* | + | | | |
| *Research Evaluation* | | | + | |
| *Research Policy* | | | + | |
| ***Scientometrics*** | + | + | + | + |

**Table 1**: Core journal lists of Information Science according to Zhao & Strotman (2008); on the basis of the citation impact of *JASIST* and *Scientometrics* in 2008; and our selection of eight journals in this study.

In a later section, we compare this set with a similar set downloaded from the *Scopus* database. This set contained 551 articles published in 2009 for the same seven journals. The difference (of 70 articles) finds its origin in the different organization of the database. *Scopus* uses publication dates and not tape years: these articles were downloaded on January 23, 2009.[2] However, the institutional addresses are differently organized in *Scopus*. Since some of our programs carefully parse the address information, we elaborated a previously existing routine (Scop2ISI.Exe, available at http://www.leydesdorff.net/software/scop2isi) in order to make the address information in the *Scopus* data as comparable with ISI-data as possible. The current version correctly displays most of the nodes and links on the maps, but the labels may still be incomplete.

The data can be processed using BibExcel (available at www.umu.se/inforsk/bibexcel/) or ISI.Exe (at http://www.leydesdorff.net/software/isi/index.htm ). The latter routine was further refined for the purpose of this project into Cities1.Exe (at http://www.leydesdorff.net/maps/index.htm). We discuss these dedicated extensions below as they become functional to the argument.

The 1,107 addresses contain 385 unique city names and 593 unique institutional addresses.[3] These could be provided with 591 and 697 geo-coordinates, respectively, at http://www.gpsvisualizer.com/geocoder/.[4] Yahoo! was used for obtaining the coordinates.[5] With the exception of one institutional address ("Isle Man Int Business Sch, Douglas 1M2 1QB, UK"), all coordinates could automatically be retrieved.

---

[2] The search string in *Scopus* was: (PUBYEAR IS 2009 AND SRCTITLE(scientometrics)) OR (PUBYEAR IS 2009 AND SRCTITLE(journal of informetrics)) OR (PUBYEAR IS 2009 AND SRCTITLE(journal of documentation)) OR (PUBYEAR IS 2009 AND SRCTITLE(journal of information science)) OR (PUBYEAR IS 2009 AND SRCTITLE(journal of the american society for information sc*)) OR (PUBYEAR IS 2009 AND SRCTITLE(information processing AND man*)) OR (PUBYEAR IS 2009 AND SRCTITLE(information research)) AND (LIMIT-TO(DOCTYPE, “ar”) OR LIMIT-TO(DOCTYPE, “ar”) OR LIMIT-TO(DOCTYPE, “ar”) OR LIMIT-TO(DOCTYPE, “ar”) OR LIMIT-TO(DOCTYPE, “ar”) OR LIMIT-TO(DOCTYPE, “ar”) OR LIMIT-TO(DOCTYPE, “ar”)) AND (LIMIT-TO(EXACTSRCTITLE, “Journal of the American Society for Information Science and Technology”) OR LIMIT-TO(EXACTSRCTITLE, “Scientometrics”) OR LIMIT-TO(EXACTSRCTITLE, “Information Processing and Management”) OR LIMIT-TO(EXACTSRCTITLE, “Journal of Information Science”) OR LIMIT-TO(EXACTSRCTITLE, “Journal of Documentation”) OR LIMIT-TO(EXACTSRCTITLE, “Journal of Informetrics”) OR LIMIT-TO(EXACTSRCTITLE, “Information Research”)).

[3] The 551 records from *Scopus* contain 801 institutional addresses of 1014 authors. At the level of the set, 495 city names are unique and 687 institutional names in this case.

[4] This number is larger because of different postcodes provided for the same city.

A co-occurrence matrix among these 392 cities (and 599 institutions, respectively) was the input to the further analysis and mapping. The city or institutional nodes can be scaled with the respective number of occurrences (or the logarithm thereof) as will be indicated where appropriate in the text. The width of the links is set proportionate to the number of co-occurrence relations. We first develop the argument with the city names because the institutional addresses generate some further complications (which will be discussed in a later section).

## 3. Google Earth and Chaomei Chen's *CiteSpace*

The first application that made it possible to generate geographic maps of science in the Google format was Chaomei Chen's program *CiteSpace*. Chen and colleagues reported about such themes such as "(v)isualizing and tracking the growth of competing paradigms" (Chen *et al*., 2002; cf. Chen, 2003) since the early 2000s. The program has been elaborated ever since and is publicly available at http://cluster.ischool.drexel.edu/~cchen/citespace/.[6] This program requires as input a download of the data in the standard (tagged) format at the Web-of-Science interface of the *Science Citation Indices* and then allows the user to make a geographic mapping of the institutional addresses and their relations—in addition to the many other facilities for citation analysis that this program offers.[7]

[5] We found the geo-encoder of Yahoo! currently more successful in retrieving Asian addresses than the one at Google.
[6] *CiteSpace* assumes the presence of the Java VM at the local computer.

When one clicks within *CiteSpace* on the tab “Geospatial Maps” in the main menu, one can select the option “Google Earth (KML)” independently of performing a citation analysis of the data. In the resulting file, links can optionally be included in addition to the nodes. The program then generates a so-called .kmz file which is the standard input for Google Earth (Figure 1).

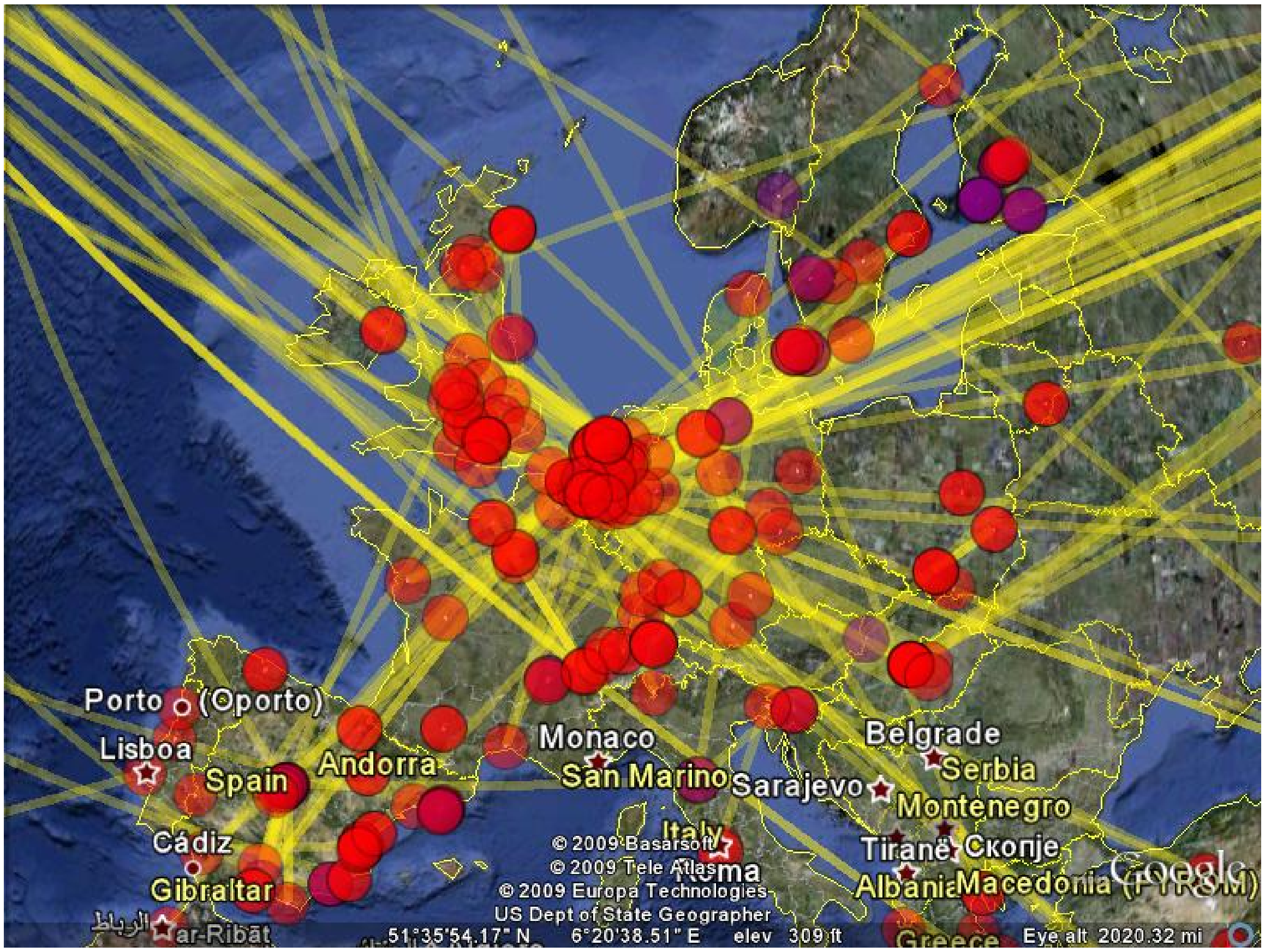

**Figure 1**: European centers and their network in the IS set 2009 as output of *CiteSpace*. (In order to enhance the visibility when printing in black and white, the color or the network links was changed from red to yellow.)[8]

[7] One can use *Scopus* data in *CiteSpace* after parsing them with Scop2Isi.Exe. The network and nodes will be correctly displayed, but the labels may sometimes contain not sufficiently standardized information.

Figure 1 provides the result for Europe using our data set. Within Google Earth, one can zoom in or out and click on links and nodes to obtain precise address information. In the output file of *CiteSpace,* the nodes vary in size as stacked bars which can be seen by tilting the image horizontally. However, in Google Earth the background is not adjustable from the satellite image to a street map like under Google Maps and because of the satellite position in the projection one is not able to draw a global map.

One therefore may wish to bring this information under Google Maps. Google Maps reads .kmz files when uploaded to a website. It is also possible to unzip .kmz-files to the .kml format which one can read and edit as a text file.[9] (KML is a markup language like HTML.) The resulting kml-file (available at http://www.leydesdorff.net/maps/master-medium.kml) contains all the information in the map, but this file cannot easily be parsed and changed, for example, in order to modify the node-sizes in accordance to the volume of publications. However, one can read this file using the web address of the upload within Google Maps. Alternatively, there are sites on the Internet where one can interactively visualize one's kml-files, such as at http://display-kml.appspot.com/. Using Google Maps, the problem of different backgrounds can be solved and one is also able to draw the global map.

---

[8] The line colors in the output of *CiteSpace* indicate different years of publications. In this case, however, the publications were all in 2009.
[9] Google Maps and Google Earth are able to read both .kmz and .kml files.

## 4. Google Maps and Google Earth

The facility to read .kml files into Google Maps provides us with many options to generate maps from the data by parsing and reformatting them into this rich markup language. However, the kml-language was primarily developed for Google Earth. (A subset of kml can also be read by Google Maps for Mobile.) Thus, the functionality in Google Maps is restricted to only a subset of tags. For example, one cannot scale the node sizes in Google Maps, but one can by using the same file in Google Earth. Google Earth, however, does not allow us to show the global map at a single glance because of the globe format of the visualization, and has the noted disadvantage of only a single "satellite view" for the mapping. However, this image can be overlayed with street names and one can tilt the image.

The various possibilities using a kml-file make this option a potentially attractive alternative for a number of applications. The zoom-facilities in Google Maps and Google Earth are superior. Thus, we decided to further develop this interface. For this purpose, the existing routine ISI.Exe[10] was further elaborated into Cities1.Exe which can be retrieved from http://www.leydesdorff.net/maps/index.htm. This program is called Cities1.Exe because after an intermediate step one needs Cities2.Exe for completing the routine. Cities1.Exe reads the same data as *CiteSpace*, but allows users to set relevant thresholds (either in absolute values or as percentages) on the fly, and to choose for

[10] Available at http://www.leydesdorff.net/software/isi/index.htm.

including the generation of a cosine normalized matrix in addition to a co-occurrence matrix.[11]

The next (and intermediate step) is reading the file cities.txt—which is one of the outputs of Cities1.Exe—at a geo-coding website which adds the geographical coordinates to the city names and postcodes. Geo-coding this information can be done, for example, at http://www.gpsvisualizer.com/geocoder/.[12] The program Cities2.Exe reads the output of this geocoder and generates, among other things, the file cities.kml (available at http://www.leydesdorff.net/maps/cities.kml) which can be uploaded and read by Google Maps or directly into Google Earth.

[11] In the case of large matrices, the generation of a cosine-normalized matrix may be time consuming. The generation of a co-occurrence matrix can be short-cutted by using the file matrix.txt in Pajek for the generation of an affiliations matrix. (This option is further explained at http://www.leydesdorff.net/maps/index.htm.)

[12] Unlike the geo-coder at Google, the one at Yahoo! led in our data to a retrieval of close to 100% of the addresses.

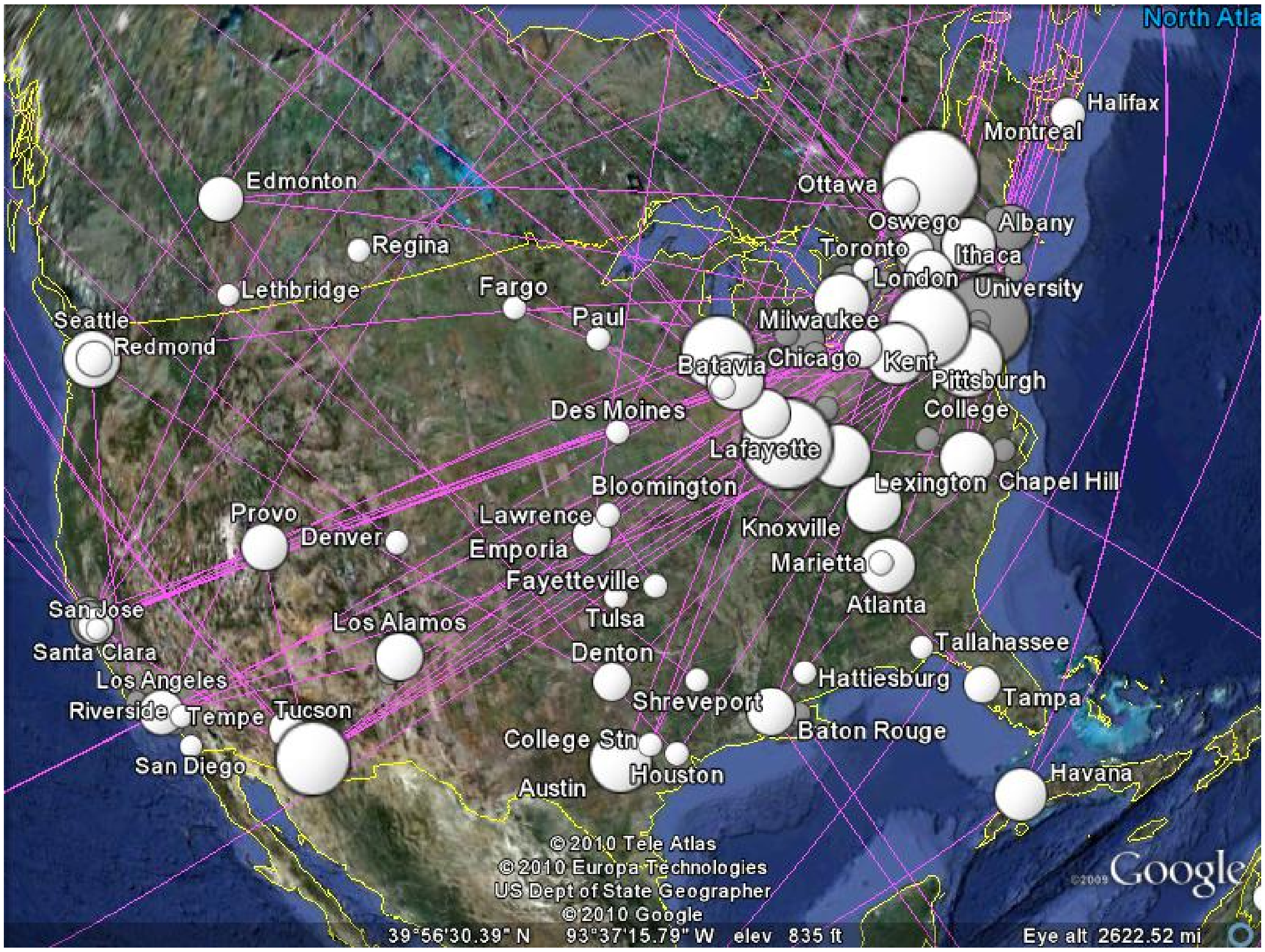

**Figure 2**: A zoom of cities.kml in Google Earth for the USA and parts of Canada.

Figure 2 shows the result in Google Earth for the United States. Using the same file in Google Maps (at http://www.leydesdorff.net/maps/cities.kml) leads to a visually awkward result because the nodes are relatively large and not scalable. This can be somewhat repaired by using a transparent icon (as at http://www.leydesdorff.net/maps/cities2.kml), but this change leads unfortunately to a systematic shift in the positioning of the cities under Google Earth.[13] However, the resulting picture becomes interesting in Google Maps because both nodes and links can be visualized, and at variable scales (e.g., globally, nationally or regionally).

[13] In some cases, we found the labeling of the links in Google Earth not reliable, while it was always in Google Maps.

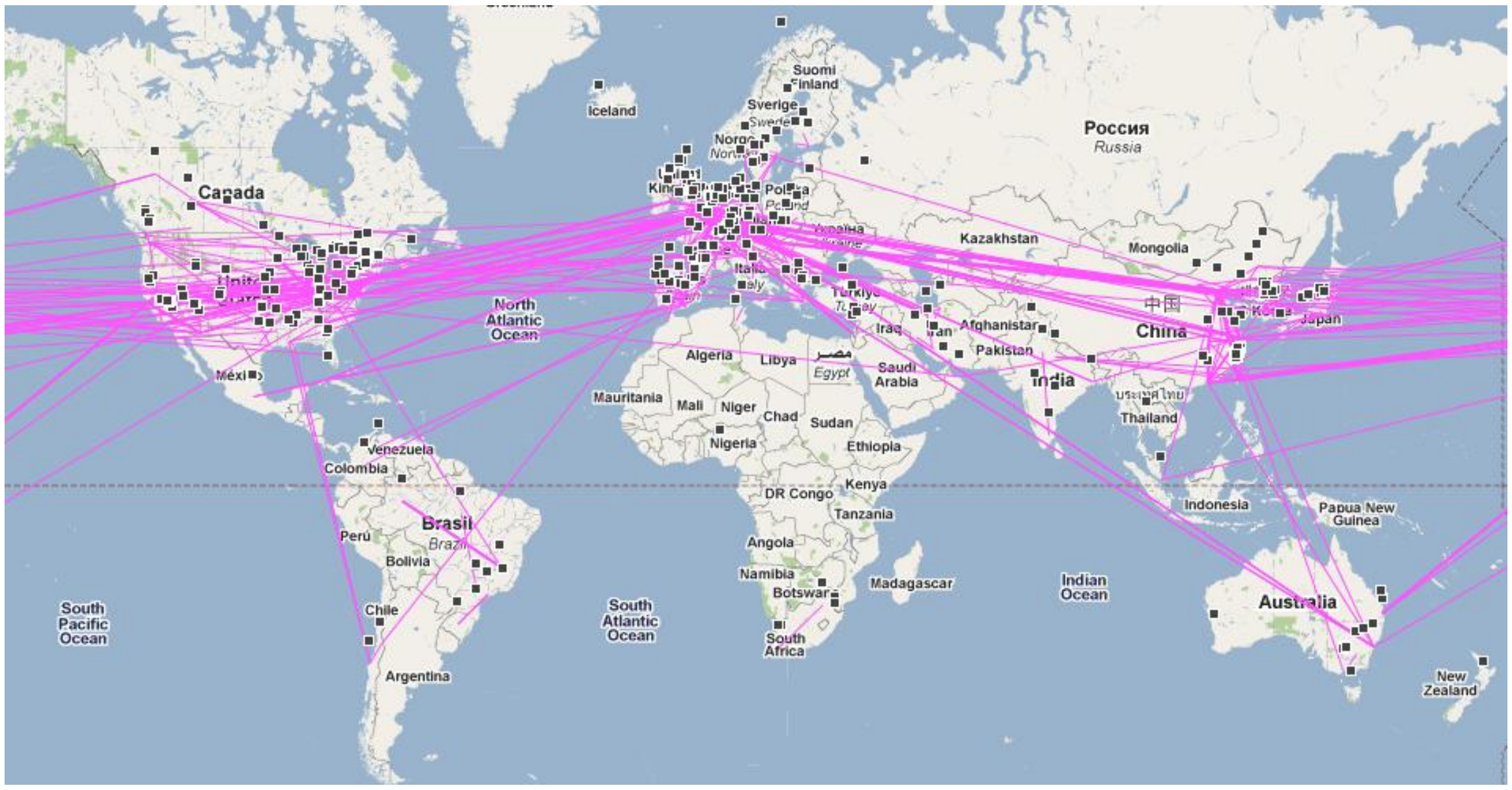

**Figure 3**: Global map of information science with the network of coauthorship relations using Google Maps (with http://www.leydesdorff.net/maps/cities2.kml).

Figure 3 shows the global map of IS in 2009 using this latter option in Google Maps. At the web, the file is clickable and zoomable. Furthermore, the user can edit the (well-structured) kml file and add information to the descriptors of nodes and links. One can also adapt the color of the links. Consequently, this file can be particularly useful for depicting network dynamics at the web (at various scales). For a dynamic animation one can collect the output of subsequent representations, for example, in a gif animator.[14]

In summary, the advantages and disadvantages of using Google Earth and/or Google Maps are a bit complex, but the kml-file offers a set of options. Google Maps is particularly useful for the global perspective and for showing the network dynamics. If the size of the vertices matters and the perspective is not global but local (or regional),

[14] The old MicroSoft GIF Animator is available as freeware, for example, at http://download.cnet.com/Microsoft-GIF-Animator/3000-18512_4-12053.html,

Google Earth provides an alternative to Google Maps since this program allows for the visualization of the sizes of the nodes.

**5. The GPS Visualizer**

As noted, the kml-language is not central to Google Maps since it was developed for Google Earth. The focus of developers is nowadays on feeding Google Maps with Javascripts using an API (that is, an application programming interface). However, this is not easy for the unskilled programmer. Fortunately, a number of websites come to the rescue of the user. One of them is the GPS Visualizer at http://www.gpsvisualizer.com/map_input?form=data. This site allows the user to input data either interactively or to read a file containing the required input information directly from one's disk. Cities2.Exe makes this file available as "inp_gps.txt." (See for an example, at http://www.leydesdorff.net/maps/inp_gps.txt.)

One can interactively change the various parameters of the data points on the Google Map to be drawn to the screen. Furthermore, the color of the nodes can be chosen in the input file (e.g., inp_gps.txt). Cities2.Exe colors connected nodes red and unconnected ones orange as the default, but one can edit the file. (Of the 392 nodes used in this study, 97 were not connected in the network.) Alternatively, BibExcel.Exe contains now a module for generating this webpage on the basis of ISI data at http://www8.umu.se/inforsk/geography/BibExcelGPSexercise.xls.

The Google Map which is generated at this interface can be saved both as a picture and in terms of the generating source code (containing Javascripts). One can adapt this source code within the html. For example, at http://www.leydesdorff.net/maps/IS2009.html, the zoom was reset at "2" instead of "1" for esthetic reasons. (Figure 4 provides a zoom of this file for East Asia.) The resulting files work promptly at one's local computer. Before the upload, however, one has to add a "Google Map API key" at the place which specifies "var google_api_key = ' ';" within the code. These API keys are freely and instantaneously available for each web address at http://code.google.com/apis/maps/signup.html.

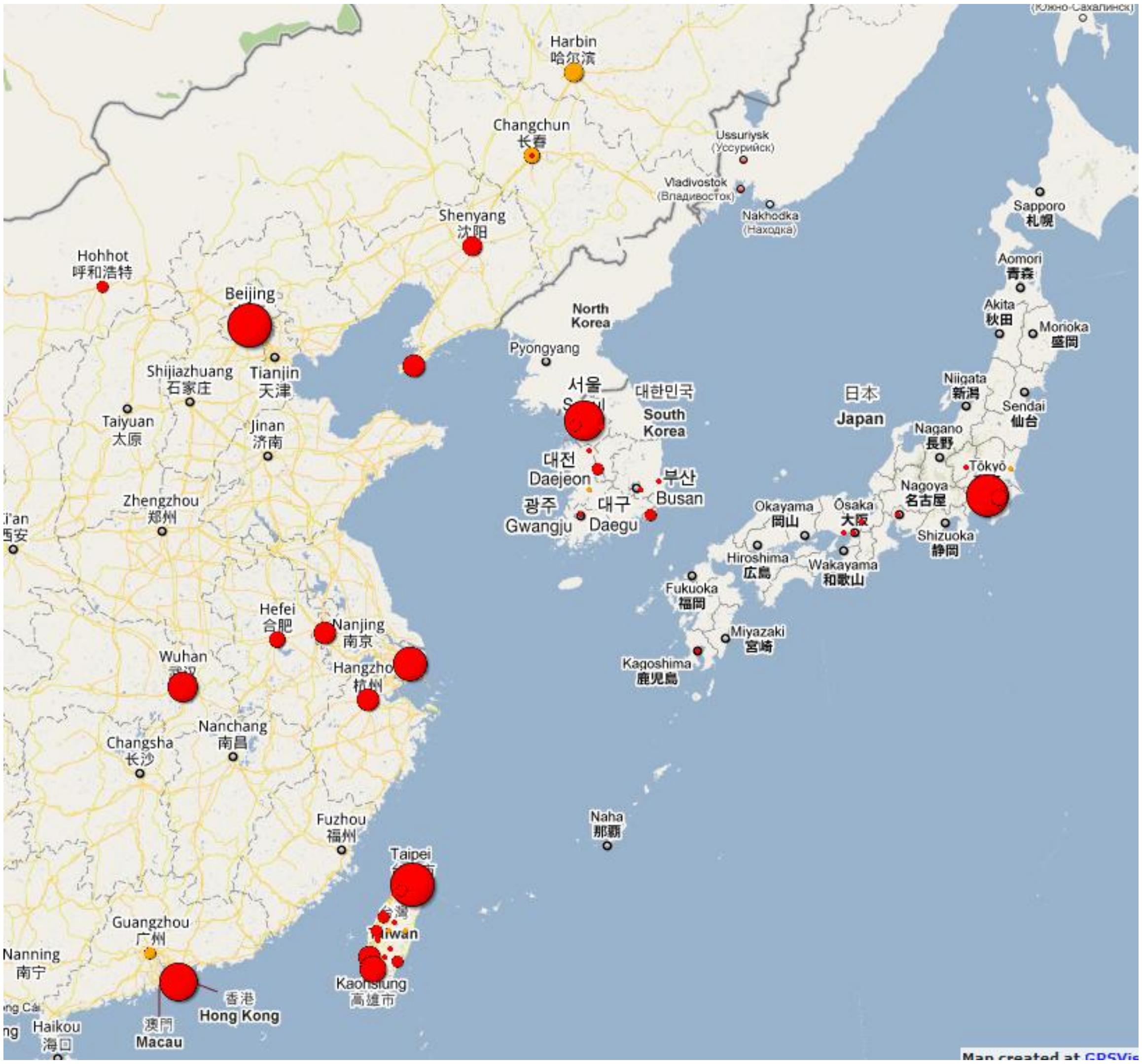

**Figure 4**: Visualization of IS in 2009 in East Asia using Google Maps via the GPS Visualizer (at http://www.leydesdorff.net/maps/is2009.html).

In summary, the use of the GPS Visualizer can have advantages above feeding kml files into Google Maps. One can vary the sizes and colors of the nodes. Furthermore, one can make an animation at the web using a so-called redirect statement in the html (e.g., <meta http-equiv="refresh" content="5;url=page2.html">).[15] However, the kml files allowed us to visualize the networks of links in addition to the nodes under Google Maps.

[15] See http://www.basictips.com/html-slideshow-5-easy-steps.shtml .

Unfortunately, one cannot have it both ways using these interfaces: one would like to be able to vary the sizes and colors of both nodes and links. Let us turn to Pajek as a network visualization program for making this possible.

**6. Pajek**

In addition to the kml files and the input for GPS Visualizer, Cities2.Exe also generates a file "cities.paj" (available at http://www.leydesdorff.net/cities.paj) which can be read into Pajek[16] as a project file (by using <F1>). Drawing this file provides a visualization with sizable arcs and vertices. The vertices are proportionate to the logarithm of the occurrences plus one (since the log(1) = 0); the links proportionate to the co-occurrences. All statistics available in Pajek can be applied (De Nooy *et al*., 2005; Hanneman & Riddle, 2005). The cities are drawn at their coordinates, and one can directly compare the geographic map with layouts generated, for example, using the algorithm of Kamada & Kawai (1989).

A layout in Pajek can be exported as a transparent overlay using the .eps format. Thus, one is able to overlay these results on any equirectangular projection of the worldmap. Additionally, we generated a worldmap in terms of coast lines which can be imported into Pajek and then merged with the overlay map.[17] This file is available in the Pajek format at http://www.leydesdorff.net/maps/coast.zip. If one reads this file into Pajek in addition to cities.paj, one obtains two networks which can both be selected (in two different Network windows) and then merged within Pajek using *Nets > Union of vertices*. One can color and size the network and the coastlines independently because the latter are defined in Pajek as edges and the former as arcs.

---

[16] Pajek is available for non-commercial use at http://vlado.fmf.uni-lj.si/pub/networks/pajek/.
[17] The coast lines are based on the geographical coordinates of the Coast Line extractor available at the website of the National Geophysical Data Center (NGDC) at http://rimmer.ngdc.noaa.gov/mgg/coast/getcoast.html. We used the World Coast Line data designed to a scale of 1:5,000,000 for this purpose. In order to match the coordinates of Pajek's Draw window, which may vary between 0 and 1, we linearly transformed the latitudes and longitudes of coastlines and cities.

One can zoom into Pajek figures by marking a piece of the drawing with a right-clicked mouse. Using the *k*-core algorithm in Pajek teaches us, for example, that the core centers of the coauthorship network in Europe are mostly in Belgium: Antwerp, Louvain,[18] Heverlee—a suburb of Louvain[19]—Oostende, and Diepenbeek. (An additional relation with Budapest is generated by Wolfgang Glänzel who adds his affiliation in Budapest routinely to his institutional address in Louvain.) Brussels moreover provides a secondary center with Hamburg, Geneva, Rouen, Paris, and Nantes. This prominent position of Belgium is an artifact of the common practice of authors in Flanders to publish papers individually at more than a single city address. We did not correct for this specific effect of the networking which is induced by policies of the regional government of Flanders (Debackere & Glänzel, 2004).[20]

[18] All 18 records in the database with "Louvain, Belgium" as address are from the Katholieke Universiteit in Leuven which publishes only a single time with its Flemish city name.
[19] One publication of the Catholic University of Leuven has exclusively an address in Heverlee, a suburb of Louvain.
[20] The Flemish government uses a model ("BOF") for the funding of basic research in academia which is based on integer counting.

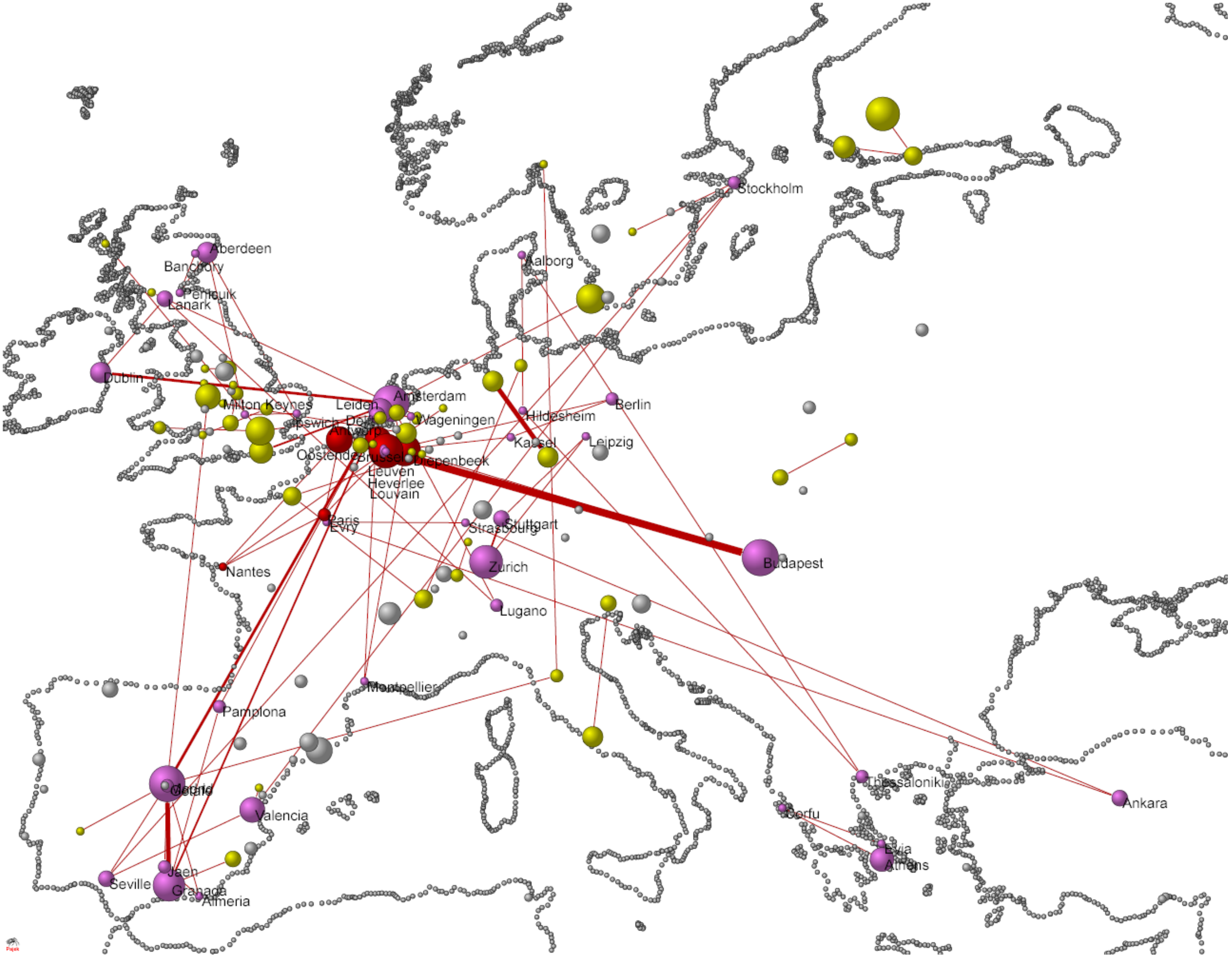

**Figure 5**: Primary and secondary centers—indicated in red and pink, respectively—in the European network of cities.

In a Pajek drawing of this network, many arcs cross the EU indicating relations between American and Asian cities. This can be prevented by choosing another projection of the earth or by refining the set (Figure 5).[1] In the exclusively European network, most core centers have lost one connection ($k = 3$ instead of $k = 4$). In this network, Budapest is no longer part of the core group of Belgian centers.

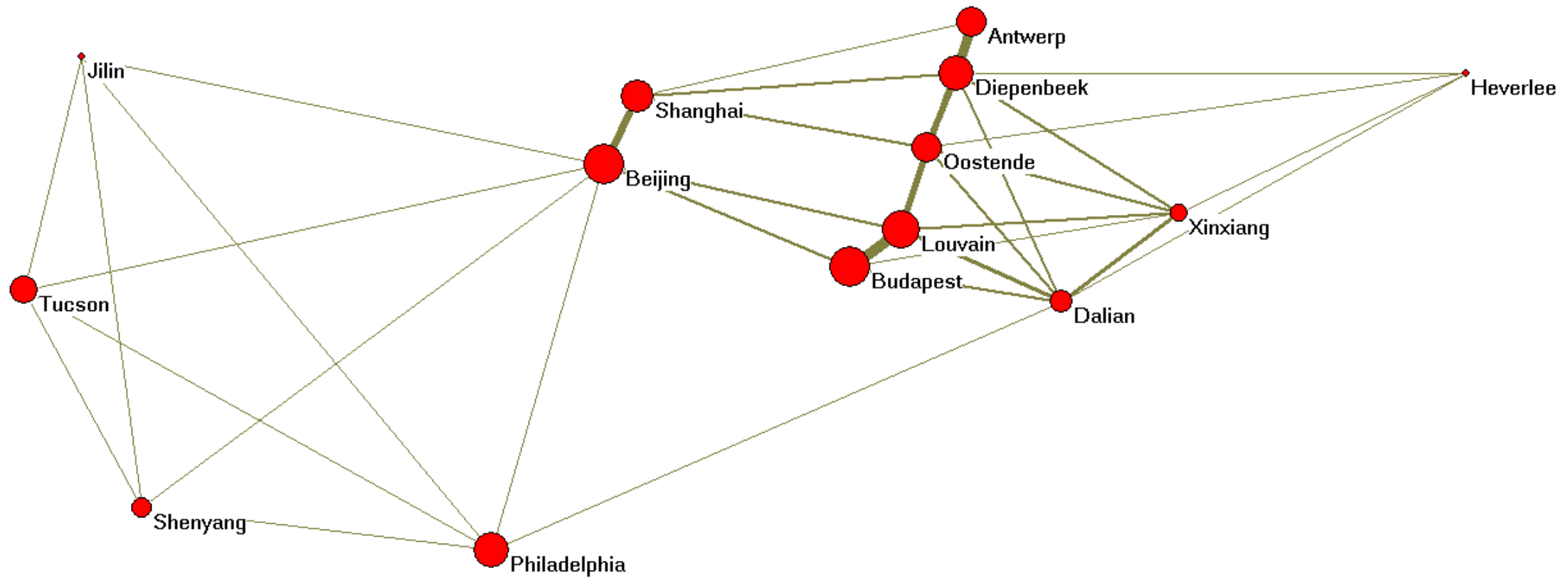

**Figure 6:** The network among 14 core cities in the network of IS 2009 (Kamada & Kawai, 1989).

Figure 6 shows the structure of the ($k \geq 4$) core network of the field. The Belgian groups do not collaborate internationally other than with cities in China and Budapest (Hungary). As noted, these relations among the Belgian (and Hungarian) cities are largely spurious. The Chinese partners have also American collaborators.

---

[1] Coast lines for Europe and Korea in Pajek-format are available at http://www.leydesdorff.net/maps/eurcoast.net and http://www.leydesdorff.net/maps/korcoast.net.

Since one can package Pajek configurations using the project file format (.paj), the information can comprehensively be communicated. (One can find the results of these analyses as examples at http://www.leydesdorff.net/maps/world.zip and http://www.leydesdorff.net/maps/europe.zip, respectively.) However, unlike Google Maps the resulting figures cannot be made interactive at the web. Animations, however, can be made using PajekToSvgAnim,[2] SoNIA[3] or the dynamic version of Visone[4] (Leydesdorff *et al*., 2008).

## 7. Institutional collaboration

Strictly analogous to the programs cities1.exe and cities2.exe, we also developed inst1.exe and inst2.exe. These latter programs include the first subfields of the institutional addresses in the ISI data in addition to the city, postcode, and country information. Using Google Maps, it thus becomes possible to map relations even at the street level.

The global map of institutions in the IS 2009 set can be retrieved at http://www.leydesdorff.net/maps/institutions.html. As before this map contains only the nodes and not the links. Of the 593 institutions, 128 centers were not connected to another one and therefore colored orange in this map; 557 institution names are unique when one disregards the different street addresses. At

[2] Available at http://vlado.fmf.uni-lj.si/pub/networks/pajek/SVGanim/.
[3] http://www.stanford.edu/group/sonia/documentation/install.html.
[4] http://www.leydesdorff.net/visone/index.htm.

http://www.leydesdorff.net/maps/inst.kml one finds the file which can be read into Google Maps and Google Earth in order to show the network relations.

There are a number of problems because the same institution may publish with different addresses and addresses are often incomplete. Costas & Irribaren-Maestro (2007) noted that valuable address information can also be found in the address of the corresponding author when it fails otherwise in the record. We include this information in the analysis although it sometimes only contains the postal address and not the institute's name.

Institutional addresses are hierarchically organized in the ISI databases with first the organization and then the sub-organization (department or faculty) after a comma as a second subfield. If the name or the organization fails, however, the sub-organization moves to the first subfield. However, a computer program cannot evaluate these differences. Thus, we used the first subfield, but always in combination with the city and country names.

Some organizations are dispersed over various address (as the Catholic University of Louvain mentioned above), but in other cases these different addresses host relatively independent organizations. For example, the *Consejo Superior de Investigaciones Científicas* (CSIC) is housed at different locations in Madrid, but also elsewhere in Spain. In our data, we found 19 records with addresses in Valencia, Sevilla, and Burgassot. In Valencia, however, this same abbreviation (CSIC) is subsumed under the Universita Polytechnica of Valencia.

In summary, the different addresses can be meaningful or not, and this cannot be decided automatically, but depends on the research question. The program Inst2.Exe therefore offers the option not to aggregate into a single institutional name. For most purposes, however, the results of these fine-grained analyses may contain considerable error. Furthermore, inst2.exe currently cannot distinguish between different locations of the same institutional name in terms of the network links. This can be further improved in the future by incorporating also into inst1.exe the option to disaggregate single institutional names in terms of different street addresses.

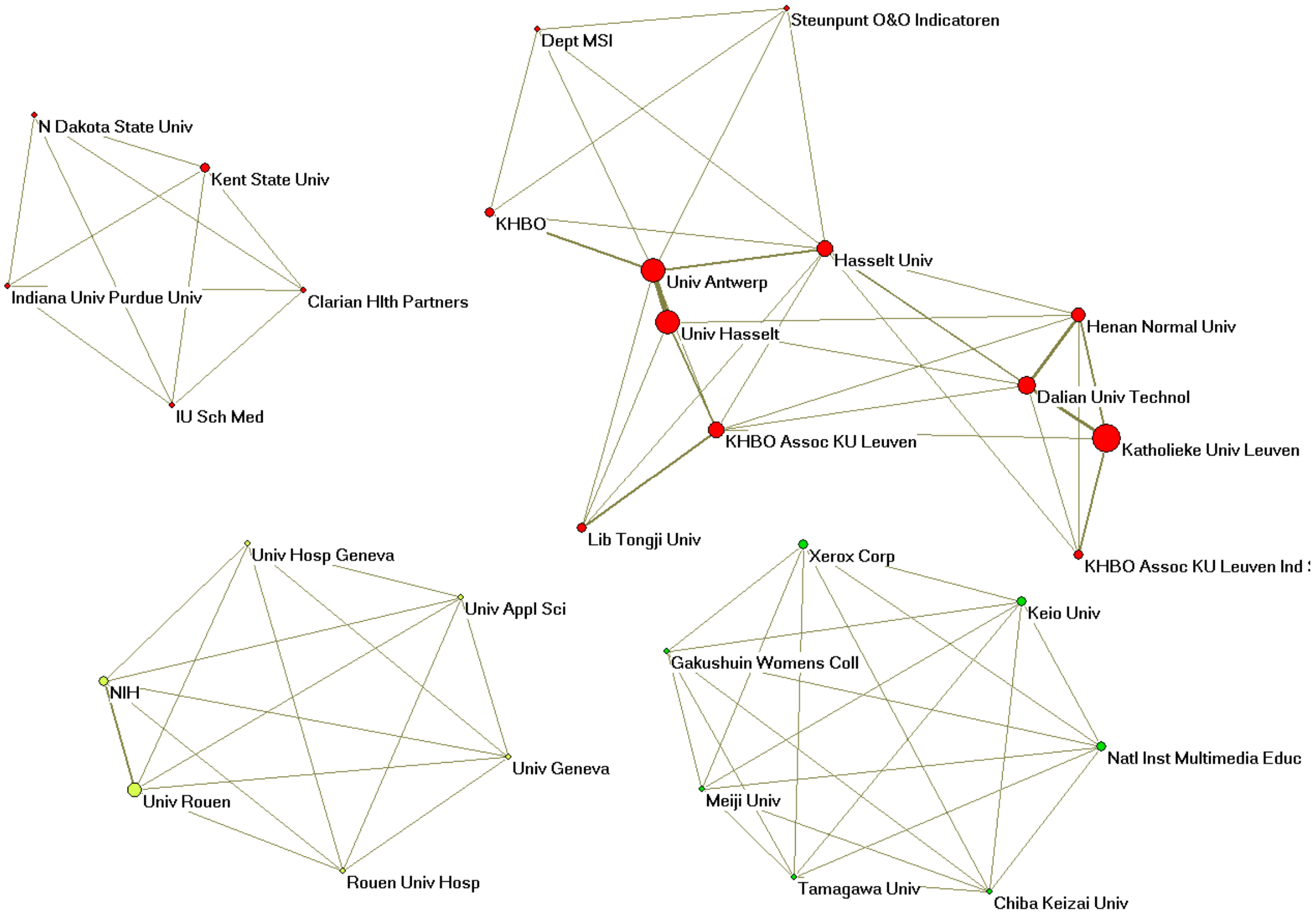

**Figure 7**: *k* > 4 networks of collaboration between leading institutes in the field of information science in 2009.

The (in this case, aggregated) institutional names provide us with a different view on the core network among these centers than was above achieved in terms of city names. Figure 7 shows a highly connected network ($k = 6$ among 7 partners) of Japanese centers and the Xerox corporation. The figure illustrates the problem of the various institutional names in the Belgian/Chinese network at the top ($k = 4$). The size of the nodes is again proportionate to the logarithm of the number of papers plus one (in order to prevent a zero as the evaluation of log(1)). Figure 7 demonstrates the effects of the noted policies of the Flemish government and the lack of standardization in the naming of institutions.

## 8. *Scopus* data

The problems with the institutional identification made us turn to *Scopus* for the comparison. Unlike the ISI databases, *Scopus* is based on index keys and one might hope that this would make a difference for the standardization. However, in this database institutional names are even less standardized than in the ISI data: even delimiters are sometimes missing. Using the 551 articles which could be retrieved with the equivalent search string, we found, for example, the three name variants "KU Leuven," "KULeuven," and "Katholieke Universiteit Leuven" among twenty records.[5] As city names, these records in *Scopus* contained "Leuven," "Leuven (Heverlee)", and "B-3000" that is, the postcode without mentioning the city. More seriously, two nodes in the Belgian network (Dalian and Xinxiang) were attributed to addresses in Taiwan according to this database (Liang & Rousseau, 2009). The geo-coder, however, recognizes this as a mistake and was able to make the correction automatically.

[5] As city names, these records in *Scopus* contained "Leuven," "Leuven (Heverlee)", and "B-3000"—that is, the postcode without mentioning the city. Thus, making the present routines fit for *Scopus* data would require another round of careful parsing of this data.

Nevertheless, the city networks using *Scopus* data are highly comparable with those based on the ISI set. These files are available at http://www.leydesdorff.net/maps/scopus.kml for Google Earth, http://www.leydesdorff.net/maps/scopus2.kml for Google Maps, http://www.leydesdorff.net/maps/scopus.html using the GPS Visualizer, and http://www.leydesdorff.net/maps/scopus.paj for Pajek. The institutional networks suffer from the same problems with inconsistent naming by authors which hitherto is beyond control for the database providers, and therefore a fortiori for users without building extensive thesauri.

## 9. Conclusions and discussion

We have wished to show the current possibilities that the bibliometric researcher can use for the visualization of one's geographic data, and hopefully provided some help by developing dedicated software to bridge existing gaps between using on the one side databases like the *Science Citation Index* and *Scopus* and on the other side the geographical projections in Google Earth, Google Map, and Pajek. (The various processing steps are summarized in an Appendix.) It seems to us that for scholarly purposes, the options in Pajek are very rich and sufficiently beautiful for the illustration. Furthermore, the data in the Pajek format can be read into a large number of available software programs; for example, at the Network Bench of Indiana University (at http://nwb.slis.indiana.edu/). Interfaces with animation programs—for time-series—are also available.

At the Google interfaces, one can import the complete dataset (as kml or kmz-file) into Google Earth, but the limitations are inherent to the satellite projection. Thus, one cannot draw the global map and one has no access to the street map. The same files can be read into Google Map. In that case, one has the full scale of projections and the network, but the nodes cannot be scaled. Using GPS Visualizer, one can scale the nodes, but one looses the network.

Which one of these options one wishes to use, depends of course on one's research question. This contribution was primarily methodological. In addition to network analysis, one can think, for example, of studies about diffusion and about correlations between distances and relations (Andersson & Persson, 1993; Katz, 1994; Wuchty *et al*., 2007). Geo-coordinates can be translated into distances using, for example, the calculator available at www.gpswaypoints.co.za/downloads/distcalc.xls.

The case in this study was selected so that the results would be recognizable in terms of flaws by this community. For example, further standardization of the address information in the bylines seems highly desirable, particularly at the institutional level. City names are currently sufficiently standardized (because of postcodes) for research purposes.

The results further clarify that co-authorship, co-location, collaboration, etc., are all different dimensions in the scientific enterprise that may or may not overlap (Katz & Martin, 1997; Wagner, 2008). The relatively new tendency to add more than a single university address to each author (Persson *et al*., 2004) further complicates the issue as we showed for the Belgian case. By making these tools available, we hope to encourage other information scientists to be able to use them in a further proliferation of research questions.

## Acknowledgement

We are grateful to Wouter de Nooy and Chaomei Chen for advice and suggestions.

**Appendix 1**: Overview of routines for the data processing.

| | | | | |
|---|---|---|---|---|
| *Scopus* data<br>↑ →<br>Scop2Isi.Exe[28] | ***Web-of-Science* data (in tagged format)** | **Output** | | |
| | | (1)<br>**kml**-files | (2)<br>**html** | (3)<br>**paj**-files |
| | →<br>→ | Cities1.Exe → Cities.txt → Geo-coding[29] → Cities2.Exe<br>Inst1.Exe → Inst.txt → Inst2.Exe | Cities.kml<br>Cities2.kml<br>Inst.kml | Inp_gps.txt | Cities.paj<br>Inst.paj |
| | | ↓<br>Matrix.txt → Possible shortcut to make co-occurrence matrix in Pajek<br>↓<br>Paj2Cooc.Exe | ↓<br>1. Use with **Google Earth**<br>2. Upload for **Google Maps**<br>3. Use at http://display-kml.appspot.com/ | ↓<br>Input to **GPS Visualizer**[30]<br>Edit the html (api-key) | ↓<br>Merge with Coast.net[31] within **Pajek** |

---

[28] Available at http://www.leydesdorff.net/software/scop2isi
[29] Available at http://www.gpsvisualizer.com/geocoder/
[30] Available at http://www.gpsvisualizer.com/map_input?form=data
[31] Available at http://www.leydesdorff.net/maps/coast.zip